\documentclass[aps,twocolumn,showpacs,preprintnumbers,prb]{revtex4}
\usepackage{graphicx}% Include figure files
\usepackage{dcolumn}% Align table columns on decimal point
\usepackage{bm}% bold math        

\begin{document}

%\preprint{paper.tex}

\title{Surface plasmons in metallic structures}
\author{J. M. Pitarke,$^{1,2}$, V. M. Silkin,$^{2}$ E. V. Chulkov,$^{2,3}$, and
P. M. Echenique$^{2,3}$}
\affiliation{
$^1$Materia Kondentsatuaren Fisika Saila, Zientzi Fakultatea,
Euskal Herriko Unibertsitatea,\\
644 Posta kutxatila, E-48080 Bilbo, Basque Country, Spain\\
$^2$Donostia International Physics Center (DIPC) and Centro Mixto
CSIC-UPV/EHU,\\ 
Manuel de Lardizabal Pasealekua, E-20018 Donostia, Basque Country, Spain\\
$^3$Materialen Fisika Saila, Kimika Fakultatea, Euskal Herriko
Unibertsitatea,\\ 
1072 Posta kutxatila, E-20080 Donostia, Basque Country, Spain}

\date{\today}

\begin{abstract}
Since the concept of a surface collective excitation was first introduced by
Ritchie,
surface plasmons have played a significant role in a variety of areas of
fundamental
and applied research, from surface dynamics to surface-plasmon microscopy,
surface-plasmon resonance technology, and a wide range of photonic
applications. Here
we review the basic concepts underlying the existence of surface plasmons in
metallic structures, and introduce a new low-energy surface collective
excitation that has been recently predicted to exist.
\end{abstract}


\maketitle

\section{Introduction}

The long-range nature of the Coulomb interaction between valence electrons in
metals is known to yield collective behaviour, manifesting itself in the form
of plasma oscillations. Pines and Bohm\cite{pines1} were the first to suggest
that the discrete energy losses experienced by fast electrons in passing
through metals are due to the excitation of these plasma oscillations,
the basic unit of energy being termed the plasmon:\cite{pines2}
$\hbar\omega_p=\hbar(4\pi ne^2/m_e)^{1/2}$, where $n$ is the valence electron
density and $m_e$ is the free electron mass.\cite{optical}

Gabor\cite{gabor} investigated the excitation of plasma oscillations in thin
foils, but assumed that the electric field is always zero at the surface. As a
result, he did not find surface modes in the bounded plasma and reached the
erroneus conclusion that the probability for plasma loss should decrease
strongly with decreasing foil thickness. Ritchie\cite{ritchie1} was the first
to find that the effect of the film boundaries is to cause the appearence of a
new "lowered" loss at $\hbar\omega_s=\hbar\omega_p/\sqrt{2}$ due to the
excitation of surface collective oscillations, the quanta of which Stern and
Ferrell called the surface plasmon.\cite{stern1} 

Ritchie's prediction of surface polarization causing low-energy losses in
metals was confirmed in a series of experiments carried out by Powell and
Swan\cite{powell}, who observed inelastic losses experienced by electrons
scattered from newly evaporated layers of Al and Mg. Since then, there has
been a significant advance in both theoretical and experimental investigations
of collective modes in the vacuum-solid interface.

The concept of the surface plasmon has played a key role in the understanding
of fundamental properties of solids and in the interpretation of a large
variety of experiments. For example, the classical image potential acting
between a point classical charge and a metal surface was shown to be
originated in the shifted zero point energy of the surface plasmon
field,\cite{feibelman0,ritchie2,mahan1,lucas1} the impact of the surfce plasmon
on surface energies was addressed,\cite{lucas2} the energy loss of charged
particles moving outside a metal surface was shown to be due to the
excitation of surface plasmons,\cite{pendry1,inkson} and the centroid of the
electron density induced by external potentials acting on a metal surface was
demonstrated to be dictated by the wave-vector dependence of surface
plasmons.\cite{feibelman} Explicit expressions for the surface-plasmon
dispersion relied originally on simple models, such as the
hydrodynamic,\cite{ritchie3} specular-reflection,\cite{ritchie4} and
infinite-barrier\cite{schaich1} models. Accurate numerical calculations have
also been performed from the knowledge of the eigenfunctions and eigenvalues of
the Kohn-Sham hamiltonian of density-functional theory (DFT),\cite{dft}
showing nice agreement with the experiments that have been carried out on
clean, well characterized surfaces of the alkali
metals.\cite{plummer2,plummer3} These experiments also showed evidence for the
existence of the so-called multipole surface plasmons that had been predicted
by Benett.\cite{benett}

The long-wavelength\cite{long} surface-plasmon energy $\hbar\omega_s$ was derived by
Ritchie in the nonrelativistic approximation, by assuming that the Coulomb
interaction is instantaneous. However, if one is to describe the interaction
of either relativistic electrons or light with solid surfaces, it is necessary
to take into account the time needed for the propagation of the true retarded
interaction. As a result of retardation, surface plasmons couple at
wavelengths larger than $\sim 2\pi c/\omega_s$ with the free
electromagnetic field and yield what is now called a surface-plasmon
polariton.\cite{ritchie5} At these large wavelengths, the surface-plasmon
polariton exists over the entire frequency range from zero to an asymptotic
value determined by the surface-plasmon frequency $\omega_s$. However, the
corresponding dispersion curve never crosses the dispersion curve of
free-space electromagnetic radiation. Hence, there is always a momentum
mismatch between light and surface plasmons of the same frequency, so that
light incident on an ideal surface cannot excite surface plasmons and,
conversely, the surface plasmon cannot decay by emitting a photon.

Teng and Stern\cite{stern2} were the first to point out that any surface
roughness permits the surface to impart some additional momentum to the
surface-plasma oscillation, so that it can couple to electromagnetic radiation.
Alternatively, prism coupling can be used to enhance the momentum of incident
light, as demonstrated by Otto\cite{otto} and by Kretchmann and
Raether.\cite{raether} Hence, surface
plasmons have been employed in a wide spectrum of studies from
electrochemistry, catalysis, wetting, thin organic condensates, and
biosensing,\cite{sambles} to scanning tunneling microscopy,\cite{stm} the
ejection of ions from surfaces,\cite{compton} surface dynamics,\cite{rocca}
surface-plasmon microscopy,\cite{nature1} and surface-plasmon resonance
technology.\cite{nature2} Moreover, recent advances that allow metals to be
structured and characterized on the nanometer scale have rekindled the
long-standing interest in surface plasmons, one of the most attractive aspects
of these collective excitations now being their use to concentrate light in
subwavelength structures\cite{pendry2} and to enhance transmission through
periodic arrays of  subwavelengths holes in optically thick metallic
films,\cite{ebbesen1,ebbesen3} as well as the possible fabrication of nanoscale
photonic circuits operating at optical frequencies.\cite{ebbesen2}

Since the typical energy of bulk and surface plasmons is of a few
electronvolts, thermal excitation of these collective oscillations is
improbable, so that the electronic properties near the Fermi level cannot be
directly influenced by these excitations. Much more effective than ordinary
bulk and surface plasmons in mediating, e.g., superconductivity would be the
so-called acoustic plasmons with sound-like long-wavelength
dispersion,\cite{pinesa} which
have spurred over the years a remarkable interest and research
activity.\cite{tosi} Acoustic plasma oscillations were observed in
two-dimensionally confined and spatially separated multicomponent structures
such as quantum wells and heterojunctions,\cite{sarma,olego} and were then
proposed as possible candidates to mediate the attractive interaction leading
to the formation of Cooper pairs in high-$T_c$
superconductors.\cite{ruvalds,kresin}

Recently, it has been shown that metal surfaces where a partially occupied
quasi two-dimensional (2D) surface-state band coexists in the same region of
space with the underlying three-dimensional (3D) continuum support a
well-defined acoustic surface plasmon.\cite{silkin} This {\it new} low-energy
collective excitation exhibits linear dispersion at low wave vectors, and
might therefore affect electron-hole (e-h) and phonon dynamics near the Fermi
level.\cite{note1} It has been demonstrated that it is a combination of the
nonlocality of the 3D dynamical screening and the spill out of the 3D electron
density into the vacuum which allows the formation of 2D electron-density
acoustic oscillations at metal surfaces, since these oscillations would
otherwise be completely screened by the surrounding 3D
substrate.\cite{pitarke1}

In this paper, we first present an overview with the basic concepts underlying
the existence of surface collective excitations in metallic structures, and we
then introduce the new concept of acoustic surface plasmons. We begin in
Section II with a brief discussion of the role that surface-plasmon excitation
plays in the interaction of fast charged particles with solid surfaces, since
it is precisely the investigation of electron energy loss in thin foils which
brought Ritchie to the realization that surface collective excitations exist
at the lowered frequency $\omega_s$.\cite{ritchie1} The detection of surface
plasmons and their dispersion is discussed in Section III, in the framework of
angle-resolved inelastic electron scattering experiments. Localized surface
plasmons and the use of sum rules that provide insight of surface-plasmon
energies in metallic structures of arbitrary geometry are introduced in Section IV.
Section V is devoted to the retarded region, where surface plasmons couple
with the free electromagnetic field. Acoustic surface plasmons are introduced
in Section VI.

Unless stated otherwise, atomic units are used throughout, i.e.,
$e^2=\hbar=m_e=1$.

\section{Plasma losses by fast charged particles in solids}

Let us consider a recoiless fast point particle of charge $Z_1$ moving in an
arbitrary inhomogeneous many-electron system with nonrelativistic velocity
${\bf v}$, for which retardation effects and radiation losses can be
neglected.\cite{notenew} The charge density of the probe particle is simply a
delta function of the form
\begin{equation}\label{rho}
\rho^{ext}({\bf r},t)=Z_1\,\delta({\bf r}-{\bf r}_0-{\bf v}\,t),
\end{equation}
and the energy that this classical particle loses per unit
time due to electronic excitations in the medium can be written
as\cite{flores}
\begin{equation}\label{flores0}
-{dE\over dt}=-\int d{\bf r}\,\rho^{ext}({\bf r},t)\,{\partial V^{ind}({\bf
r},t)\over\partial t},
\end{equation}
where $V^{ind}({\bf r},t)$ is the potential induced by the probe particle at
position ${\bf r}$ and time $t$.

To first order in the external perturbation, time-dependent perturbation theory
yields
\begin{eqnarray}\label{vi}
V^{ind}({\bf r},t)&=&\int d{\bf
r}'\int_{-\infty}^{+\infty} dt'\int_{-\infty}^{+\infty}{d\omega\over
2\pi}\,{\rm e}^{-i\omega(t-t')}\cr\cr\cr
&\times&\tilde W({\bf r},{\bf r}';\omega)\,\rho^{ext}({\bf r}',t'),
\end{eqnarray}
where
\begin{equation}
\tilde W({\bf r},{\bf r}';\omega)=W({\bf r},{\bf r}';\omega)-v({\bf r},{\bf
r}'),
\end{equation}
$v({\bf r},{\bf r}')$ being the bare Coulomb interaction and $W({\bf r},{\bf
r}';\omega)$ being the so-called screened interaction, which is
usually expresed in terms of the density-response function of the
many-electron system.\cite{fetter}

Eq.~(\ref{vi}) is a general expression for the energy loss of a classical
particle moving in an arbitrary inhomogeneous electron system that is
characterized by the screened interaction $W({\bf r},{\bf r}';\omega)$. Here
we consider a solid target consisting of a fixed uniform positive background
(jellium) plus a neutralizing cloud of interacting valence electrons, which
will be described by either an infinite or a plane-bounded electron gas.

\subsection{Infinite electron gas}

In the case of an infinite homogeneous electron gas that is translationally
invariant in all directions, Eqs.~(\ref{rho})-(\ref{vi}) yield the following
expression for the so-called stopping power, i.e., the energy
that the probe particle loses per unit path length:
\begin{equation}\label{uniform}
-{dE\over dx}=-Z_1^2\int{d^3{\bf k}\over(2\pi)^3}\,{\bf k}\cdot{\bf v}\,{\rm
Im}\tilde W(k,{\bf k}\cdot{\bf v}),
\end{equation} 
$\tilde W(k,\omega)$ being the 3D Fourier transform
of $\tilde W({\bf r},{\bf r}';\omega)$, which is typically expressed in the
form
\begin{equation}
\tilde W(k,\omega)={4\pi\over k^2}\left[\epsilon^{-1}(k,\omega)-1\right],
\end{equation}
where $\epsilon^{-1}(k,\omega)$ is the so-called inverse dielectric
function of the electron gas.

At high projectile velocities ($v_F<<v$, $v_F$ being the Fermi velocity), the
zero-point motion of the electron gas can be neglected and it can be
considered,
therefore, as if it were at
rest. In this approximation, the dielectric function $\epsilon(k,\omega)$ of a
homogeneous electron gas takes the form\cite{lindhard}
\begin{equation}\label{static}
\epsilon(k,\omega)=1-{\omega_p^2\over\omega(\omega+i\eta)-k^4/4},
\end{equation}
which at long wavelengths ($k\to 0$) yields the classical Drude dielectric
function
\begin{equation}\label{drude}
\epsilon(\omega)=1-{\omega_p^2\over\omega(\omega+i\eta)},
\end{equation}
$\omega_p$ being the bulk plasma frequency and $\eta$ a positive infinitesimal.

The dielectric function of Eq.~(\ref{static}) describes both collective and
single-particle excitations. At wave vectors ${\bf
k}$ smaller than a cut-off wave vector of magnitude $k_c\sim\omega_p/v_F$, the
many-electron system can be expected to behave collectively, losses being
dominated by the excitation of plasma oscillations.\cite{pines2} At
wave vectors of
magnitude larger than $k_c$, losses are dominated by the excitation of
electron-hole (e-h) pairs. Using Eq.~(\ref{static}) and assuming that 
$v>>v_F$,
Eq.~(\ref{uniform}) yields
\begin{equation}\label{split}
-{dE\over dx}=Z_1^2\,{\omega_p^2\over v^2}\left[{\rm
ln}{k_c\over\omega_p/v}+{\rm
ln}{2v\over k_c}\right],
\end{equation}
or, equivalently,
\begin{equation}
-{dE\over dx}=Z_1^2\,{\omega_p^2\over v^2}\,{\rm ln}{2v^2\over\omega_p}.
\end{equation}

The first term of Eq.~(\ref{split}), which can also be obtained by using the
Drude
dielectric function of Eq.~(\ref{drude}), represents the contribution to the
stopping
power from losses to collective excitations at wave vectors of magnitude
smaller than
$k_c$. The second term of Eq.~(\ref{split}) represents the contribution from
losses to   
single-particle excitations at wave vectors above $k_c$.

\subsection{Plane-bounded electron gas}

In the case of a plane-bounded electron gas that is translationally invariant
in two
directions, which we take to be normal to the $z$ axis,
Eqs.~(\ref{rho})-(\ref{vi})
yield the following expression for the energy that the probe particle loses per
unit
time:
\begin{eqnarray}\label{general}
-{dE\over dt}&=&i\,{Z_1^2\over\pi}\int {d^2{\bf
q}\over(2\pi)^2}\int_{-\infty}^{+\infty}
dt' \int_0^\infty d\omega\,\omega\cr\cr\cr
&\times&{\rm e}^{-i(\omega-{\bf
q}\cdot{\bf v}_\parallel)(t-t')}\,\tilde W[z(t),z(t');q,\omega],
\end{eqnarray}
where ${\bf q}$ is a 2D wave vector in the plane of the surface,
${\bf v}_\parallel$ represents the component of the velocity that is parallel
to the surface, $z(t)$ represents the position of the projectile relative to
the surface, and  $\tilde W(z,z';q,\omega)$ is the 2D Fourier transform
of $\tilde W({\bf r},{\bf r}';\omega)$.

In the simplest possible model of a bounded semi-infinite electron gas, one
characterizes the electron gas at $z\geq 0$ by a local dielectric function
which jumps discontinuosly at the surface from unity outside ($z<0$) to
$\epsilon(\omega)$ inside ($z>0$). By imposing the ordinary boundary
conditions of continuity of the potential and the normal component of the
displacement vector at $z=0$, one finds
\begin{widetext}
\begin{equation}\label{w}
\tilde W(z,z';{\bf q},\omega)={2\pi\over q}\cases{
-g(\omega)\,{\rm e}^{-q(|z|+|z'|)}& $z<0$\cr\cr
\left[\epsilon^{-1}(\omega)-1\right]{\rm
e}^{-q|z-z'|}+\epsilon^{-1}(\omega)\,g(\omega)\,{\rm e}^{-q(|z|+|z'|)},&
$z>0$,}
\end{equation}
\end{widetext}
where
\begin{equation}\label{g}
g(\omega)={\epsilon(\omega)-1\over \epsilon(\omega)+1}
\end{equation}
is the long-wavelength ($q\to 0$) limit of the so-called surface-response
function.\cite{liebsch} In this limit, the dielectric function
$\epsilon(\omega)$ takes
the Drude form dictated by Eq.~(\ref{drude}).

In the following, we shall explicitly consider particle trajectories that are
normal and parallel to the surface.

\subsubsection{Normal trajectory}

Let us consider a situation in which the probe particle moves along a normal
trajectory from the vacuum side of the surface ($z<0$) and enters the solid at
$z=t=0$. The position of the projectile relative to the surface is then
$z(t)=vt$. Assuming that the electron gas at $z\geq 0$ can be described by the
Drude dielectric function of Eq.~(\ref{drude}) and introducing Eq.~(\ref{w})
into Eq.~(\ref{general}), one finds the following expression for the energy
that the probe particle loses per unit path length:
\begin{widetext}
\begin{equation}\label{normal}
-{dE\over dz}={Z_1^2\over v^2}\cases{\omega_s^2\,f(2\omega_s|z|/v),&$z<0$\cr\cr
\omega_p^2\left[{\rm
ln}(k_cv/\omega_p)-h(\omega_pz/v)\right]+\omega_s^2\,h(\omega_sz/v),&$z>0$,}
\end{equation}
\end{widetext}
where
\begin{equation}\label{b2}
h(\alpha)=2\cos(\alpha)\,f(\alpha)-f(2\alpha),
\end{equation}
with $f(\alpha)$ being given by the following expression:
\begin{equation}
f(\alpha)=\int_0^\infty{x\,{\rm e}^{-\alpha x}\over 1+x^2}.
\end{equation}

The Drude dielectric function of Eq.~(\ref{drude}), which assumes that
infinitely long-lived plasmons at a single-frequency $\omega_p$ are the only
possible bulk excitations, is only sustainable at wave vectors below a cut-off
$k_c$. Hence, this cut-off has been introduced into the bulk (logarithmic) term
of Eq.~(\ref{normal}), which yields a contribution to the energy loss that
coincides with the plasmon contribution of Eq.~(\ref{split}). Contributions to
the energy loss that are due to the excitation of e-h pairs are not included
in Eq.~(\ref{normal}).

In the absence of the boundary at $z=0$, the position-dependent $f(\alpha)$ and
$h(\alpha)$ terms entering Eq.~(\ref{normal}) would not be present, and the
energy loss
would be that of charged particles moving in an infinite plasma. When the probe
particle is
moving outside the solid, the effect of the boundary is to cause energy loss at
the lowered
plasma frequency $\omega_s$. When the probe particle is moving inside the
solid, the effect
of the boundary is to cause both a decrease in loss at the bulk plasma
frequency $\omega_p$
and an additional loss at the lowered plasma frequency $\omega_s$, as predicted
by Ritchie.\cite{ritchie1}

\begin{figure}
\centering
\includegraphics[width=0.5\linewidth]{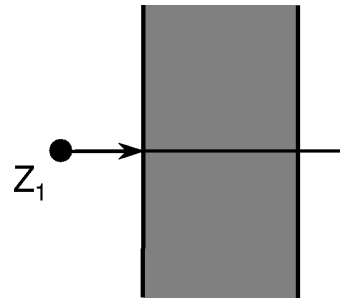}\\
\caption{Particle of charge $Z_1$ passing through a finite foil of thickness
$a$.} \label{fig1}
\end{figure}

Now we consider the real situation in which a fast charged particle passes
through a {\it finite} foil of thickness $a$ (see Fig.~\ref{fig1}). Assuming
that the foil is thick enough
for the effect of each boundary to be the same as in the case of a
semi-infinite medium, and
integrating along the whole trajectory from minus to plus infinity, one finds
the total energy
that the probe particle loses to collective excitations:
\begin{equation}\label{delta}
-\Delta E={Z_1^2\over v^2}\left[a\,\omega_p^2\,{\rm ln}{k_cv\over\omega_p}
-{\pi\over 2}\omega_p+\pi\omega_s\right].
\end{equation}

This is the result first derived by Ritchie in a different way,\cite{ritchie1}
which brought him to the realization that surface collective excitations exist
at the lowered frequency $\omega_s$. The first term of Eq.~(\ref{delta}),
which is proportional to the thickness of the film represents the bulk
contribution, which would also be present in the absence of the
boundaries. The second and third terms, which are both due to the presence of
the boundaries and become more important as the foil thickness decreases,
represent the decrease in the energy loss at the plasma frequency $\omega_p$
and the energy loss at the lowered frequency $\omega_s$, respectively.
Eq.~(\ref{delta}) also
shows that the net boundary effect is an increase in the total energy loss
above the value
which would exist in its absence, as noted by Ritchie.\cite{ritchie1}

Ritchie also considered the coupling that exists between the two surfaces for
finite values of the film thickness $a$. He found the following dispersion
relation between the frequency of surface-plasma oscillations and the wave
number $q$:
\begin{equation}\label{dispersion}
\omega=\omega_s\left[1\pm{\rm e}^{-aq}\right]^{1/2},
\end{equation}
the exponential factor being a consequence of the interaction between the two
surfaces. This equation has two limiting cases, as discussed by
Ferrell.\cite{ferrell0} At short wavelengths ($qa>>1$), the surface waves
become decoupled and each surface sustains independent oscillations at the
reduced frequency $\omega_s$  characteristic of a semi-infinite electron gas
with a single plane boundary. At long wavelengths ($qa<<1$), there are
"normal" oscillations at $\omega_p$ and "tangential" 2D oscillations at
\begin{equation}\label{2D}
\omega_{2D}=(2\pi naq)^{1/2},
\end{equation}
which were later discussed by Stern\cite{stern} and observed in artificially
structured semiconductors\cite{allen} and more recently in a metallic
surface-state band on a silicon surface.\cite{nagao}

\subsubsection{Parallel trajectory}

\begin{figure}
\includegraphics[width=0.5\linewidth]{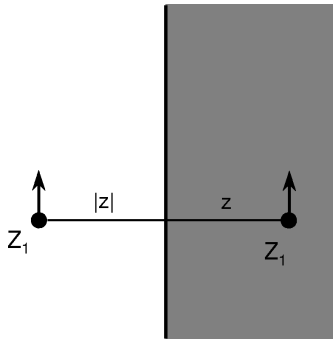}\\
\caption{Particle of charge $Z_1$ moving at a fixed distance $z$ from the
surface of a plane-bounded electron gas.}\label{fig2}
\end{figure}

Now we restrict our attention to the case of charged particles moving with
constant velocity at a fixed distance $z$ from the surface (see
Fig.~\ref{fig2}).
Eq.~(\ref{general}) then yields
\begin{eqnarray}\label{stop}
-{dE\over dx}=&&-{2\over v}\,Z_1^2\int{d^2{\bf
q}\over(2\pi)^2}\int_0^\infty d\omega\,\omega\cr\cr
&&\times{\rm
Im}\tilde W(z,z;{\bf q},\omega)\,\delta(\omega-{\bf q}\cdot{\bf
v}).
\end{eqnarray}

Assuming that the electron gas at $z\ge 0$ can be described by the
Drude dielectric function of Eq.~(\ref{drude}) and introducing Eq.~(\ref{w})
into Eq.~(\ref{stop}), one finds
\begin{widetext}
\begin{equation}\label{parallel}
-{dE\over dx}={Z_1^2\over
v^2}\cases{\omega_s^2\,K_0(2\omega_s|z|/v),&$z<0$\cr\cr
\omega_p^2\left[{\rm
ln}(k_cv/\omega_p)-K_0(2\omega_pz/v)\right]+\omega_s^2
\,K_0(2\omega_sz/v),&$z>0$,}
\end{equation}
\end{widetext}
where $K_0(\alpha)$ is the zero-order modified Bessel function.\cite{abra} 

For particle trajectories outside the solid ($z<0$), Eq.~(\ref{parallel})
reproduces the classical expression of Echenique and Pendry.\cite{pendry1} For
particle trajectories inside the solid ($z>0$), Eq.~(\ref{parallel})
reproduces the result first obtained by Nu\~nez {\it et al.}.\cite{nunez} As in
the case of a normal trajectory, when the particle moves inside the solid the
effect of the boundary is to cause a decrease in loss at the bulk plasma
frequency $\omega_p$ and an additional loss at the lowered plasma frequency
$\omega_s$.

\section{Inelastic electron scattering}

\begin{figure}
\includegraphics[width=0.5\linewidth]{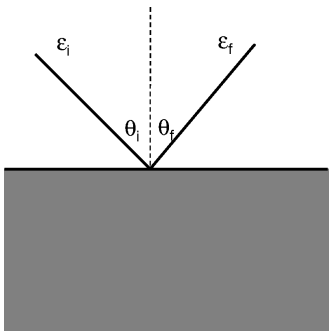}\\
\caption{Schematic drawing of the scattering geometry in angle-resolved
inelastic electron scattering experiments. By exciting a surface mode of
frequency $\omega(q)$, the energy of detected electrons becomes
$\varepsilon_f=\varepsilon_i-\omega(q)$. The momentum $q$ is determined from
$q=\sqrt{2}\left[\sqrt{\varepsilon_i}\,\sin\theta_i-\sqrt{\varepsilon_f}\,\sin\theta_f\right]$}\label{fig3}
\end{figure}

The most commonly used experimental arrangement for detecting surface plasmons
by the fields of moving charged particles is based on angle-resolved inelastic
electron scattering.\cite{ibach,plummer4} Fig.~\ref{fig3} shows a schematic
drawing of the scattering geometry. A monochromatic beam of electrons,
incident on a flat surface at an angle $\theta_i$, is back scattered and
detected by an angle-resolved energy analyzer positioned at an angle
$\theta_f$. Inelastic events can occur, either before or after the elastic
event, by exciting a surface mode of frequency $\omega(q)$. The energy and
lifetime of this mode are determined by the corresponding energy-loss peak in
the spectra, and the momentum $q$ parallel to the surface is obtained from the
measured angles $\theta_i$ and $\theta_f$.

The inelastic-scattering cross section corresponding to a process in which an
electronic excitation of energy $\omega$ and parallel momentum $q$ is created
at the surface of a semi-infinite electron gas is dictated by the imaginary
part of the surface-response function $g(q,\omega)$,\cite{liebsch} which in the
long-wavelength limit is given by Eq.~(\ref{g}). In a free-electron gas
(jellium), the dielectric function $\epsilon(\omega)$ entering Eq.~(\ref{g})
is the Drude dielectric function of Eq.~(\ref{drude}). Hence, in a gas of free
electrons ${\rm Im}g(q,\omega)$ becomes a delta function peaked at the
surface-plasmon energy $\omega_s$.

\begin{figure}
\includegraphics[width=0.5\linewidth]{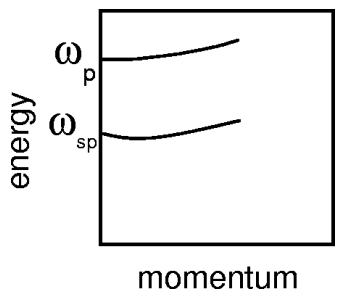}\\
\caption{Schematic drawing of the bulk and surface plasmon energy dispersions
in typical metal surfaces. While in the case of bulk plasmons the initial
slope is usually (but not always) positive, the surface-plasmon energy
dispersion is negative for typical metal surfaces.}\label{fig4}
\end{figure}

The classical picture leading to Eqs.~(\ref{drude}) and (\ref{g}), which is
correct only in the long-wavelength limit,  ignores both
the nonlocality of the electronic response of the system and the microscopic
spatial distribution of the electron density near the surface. Feibelman
showed that up to first order in an expansion in powers of $q$, the
surface-response function of a jellium surface can be written
as\cite{feibelman}
\begin{equation}\label{expansion}
g(q,\omega)={\epsilon(\omega)-1\over\epsilon(\omega)+1}\left[1+2qd_\perp(\omega)
{\epsilon(\omega)\over\epsilon(\omega)+1}\right]+O(q^2),
\end{equation}
where $d_\perp(\omega)$ represents the centroid of the induced electron
density. Self-consistent density-functional calculations of
$g(q,\omega)$ and $d_\perp(\omega)$ have demonstrated that in the case of jellium
surfaces in the range of typical bulk densities the surface-plasmon energy
$\omega(q)$ [where ${\rm Im}g(q,\omega)$ is maximum] is at nonvanishing but
small ${\bf q}$ wave vectors lower than $\omega_s$, in agreement with
experiment.\cite{plummer2,plummer3} This is illustrated in Fig.~\ref{fig4},
where the surface-plasmon energy dispersion is drawn schematically, showing
that the initial slope is negative. For an interpretation of negative
dispersion and comparison to experiment see Ref.~\onlinecite{liebsch}.

\begin{table}
\caption{Relative widths $\Delta\omega_s/\omega_s$ of surface plasmons, as
derived from the imaginary part of the surface-response function of
Eq.~(\ref{g}) with measured values of the bulk dielectric function
$\epsilon(\omega)$ (theory)\cite{liebsch} and from the surface-loss
measurements at $q=0$ (experiment).\cite{plummer5}}
\begin{ruledtabular} \begin{tabular}{lcccccc}
%&\multicolumn{3}{c}{correlation}&\multicolumn{3}{c}{exchange-correlation}\\
&Ag&K&Al&Mg&Hg&Li\\ \hline
Theory&0.027&0.035&0.035&0.16&0.18&0.33\\
Experiment&0.027&0.1&0.24&0.16&0.16&0.35\\
\end{tabular} \end{ruledtabular} \label{table1}
\vspace{0.5 cm}
\end{table}

At jellium surfaces, the actual surface-response function reduces in the
long-wavelength limit to Eq.~(\ref{g}) with the Drude $\epsilon(\omega)$ of
Eq.~(\ref{drude}), and surface plasmons are therefore expected to be
infinitely long-lived excitations. However, energy-loss measurements at simple
metal surfaces indicate that surface plasmons exhibit a finite width even at
$q=0$.\cite{plummer5} Since surface plasmons are dictated in this $q=0$ limit
by bulk properties through the dielectric function $\epsilon(\omega)$, the
experimental surface-plasmon widths $\Delta\omega_s$ at $q=0$ should be
approximately described by using in Eq.~(\ref{g}) the measured bulk dielectric
function $\epsilon(\omega)$ instead of its Drude counterpart.
Table~\ref{table1} exhibits the relative widths $\Delta\omega_s/\omega_s$ of
surface plasmons derived in this way,\cite{liebsch} together with surface-loss
measurements at $q=0$.\cite{plummer5} Although in the case of Ag, Li, Hg, and
Mg the surface-plasmon width is well described by introducing the measured
bulk dielectric function into Eq.~(\ref{g}), the surface-plasmon widths of K and Al are
considerably larger than predicted in this simple way. 
This shows that an understanding of surface-plasmon broadening mechanisms
requires a careful analysis of the actual band structure of the solid.
Approximate treatments of the impact of the band structure on the
surface-plasmon energy dispersion have been developed by several
authors,\cite{zaremba,apell,tarriba,feibelmanl,liebschl} but a
first-principles description of surface energy-loss measurements has not been
carried out yet.

\section{Sum rules}

\begin{figure}
\includegraphics[width=0.95\linewidth]{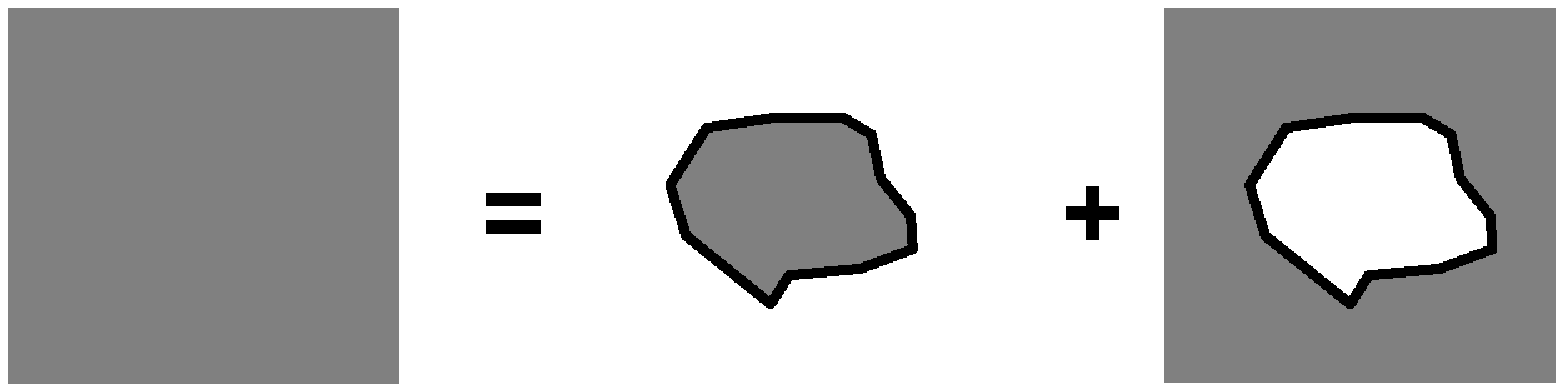}\\
\vspace{1cm}
\includegraphics[width=0.95\linewidth]{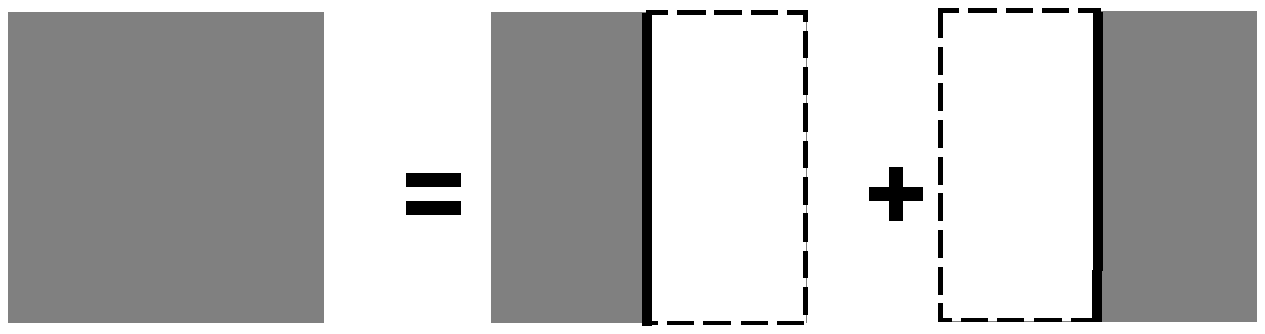}\\
\caption{Complementary systems in which the regions of plasma and vacuum are
interchanged. The top panel represents the general situation. The bottom panel
represents a half-space filled with metal and interfaced with vacuum. The
surface-mode frequencies $\omega_{s_1}$ and $\omega_{s_2}$
of these systems fulfill the sum rule of Eq.~(\ref{sum1}).}\label{fig5} 
\end{figure}

Sum rules have played a key role in providing insight in the investigation of a
variety of physical situations. A useful sum rule for the surface modes in
complementary media with arbitrary geometry was introduced by Apell {\it et
al.},\cite{apell1} which
in the special case of a metal/vacuum interface implies that\cite{ronveaux}
\begin{equation}\label{sum1}
\omega_{s_1}^2+\omega_{s_2}^2=\omega_p^2,
\end{equation}
where $\omega_{s_1}$ is the surface-mode frequency of our system, and
$\omega_{s_2}$ represents the surface mode of a second complementary system in
which the regions of plasma and vacuum are interchanged (see Fig.~\ref{fig5}).

For example, a half-space filled with a metal of bulk plasma frequency
$\omega_p$ and interfaced with vacuum maps into itself (see Fig.~\ref{fig5}),
and therefore Eq.~(\ref{sum1}) yields
\begin{equation}
\omega_{s_1}=\omega_{s_2}=\omega_p/\sqrt{2},
\end{equation}
which is the frequency of plasma oscillations at a metal/vacuum planar
interface.

Other examples are a metal sphere in vacuum, which sustains {\it localized} Mie
plasmons at frequencies
\begin{equation}\label{l1}
\omega_l=\omega_p\,\sqrt{l\over 2l+1},
\end{equation}
with $l=1,2,\dots$, and a spherical void in a metal, which shows Mie plasmons
at frequencies
\begin{equation}\label{l2}
\omega_l=\omega_p\,\sqrt{l+1\over2l+1}.
\end{equation}
The squared surface-mode frequencies of the sphere [Eq.~(\ref{l1})] and the
void [(Eq.~(\ref{l2})] add up to $\omega_p^2$ for all $l$, as required by
Eq.~(\ref{sum1}).

Now we consider a situation in which there are two interfaces, as occurs in the
case of a thin film and approximately occurs in the case of multishell
fullerenes\cite{nano1} and carbon nanotubes.\cite{nano2} Apell {\it et
al.}\cite{apell1} have proved a second sum rule, which relates the surface
modes corresponding to the in-phase and out-of-phase linear combinations of
the screening charge densities at the interfaces. In the case of metal/vacuum
interfaces this sum rule takes the form of Eq.~(\ref{sum1}), but
now $\omega_{s_1}$ and $\omega_{s_2}$ being in-phase and out-of-phase modes of
the same system.

\begin{figure}
\includegraphics[width=0.5\linewidth]{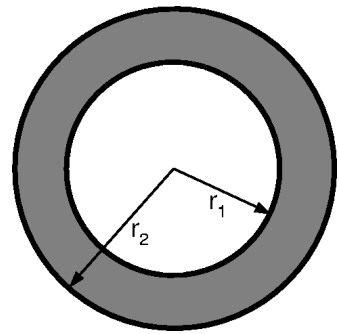}\\
\caption{Fullerene molecule of inner and outer radii $r_1$ and $r_2$. A Drude
dielectric function is assigned to every point between the inner and outer
surfaces.}\label{fig7}
\end{figure}

For a Drude metal film with equal and abrupt planar surfaces, the actual values
of $\omega_{s_1}$ and $\omega_{s_2}$ are those given by
Eq.~(\ref{dispersion}). For a spherical fullerene molecule described by
assigning a Drude dielectric function to every point between the inner and
outer surfaces of radii $r_1$ and $r_2$ (see Fig.~\ref{fig7}), one finds the
following
frequencies for the in-phase and out-of-phase surface modes:\cite{apell2}
\begin{equation}
\omega_s^2={\omega_p^2\over 2}\left[1\pm{1\over
2l+1}\sqrt{1+4l(l+1)(r_1/r_2)^{2l+1}}\right],
\end{equation}
which fulfill the sum rule dictated by Eq.~(\ref{sum1}).

\section{Surface-plasmon polaritons}

Planar surface plasmons are known to be traced to the peaks of the imaginary part of
the surface response function $g(q,\omega)$,\cite{liebsch} which in the
long-wavelength limit is given by Eq.~(\ref{g}).  This equation yields the
{\it classical} non-dispersive surface-plasmon frequency $\omega_s$ given by
\begin{equation}\label{classical}
\epsilon(\omega_s)+1=0,
\end{equation}
which in the case of a semi-infinite Drude metal [see Eq.~(\ref{drude})] is
$\omega_s=\omega_p/\sqrt 2$.

The long-wavelength surface-plasmon condition of Eq.~(\ref{classical}) has been
derived in the nonrelativistic approximation, by neglecting retardation of the
Coulomb interaction. Hence, Eq.~(\ref{classical}) yields a good representation
of surface plasma oscillations only at  wavelengths that are large compared to
the Fermi wavelength ($q<<k_F\sim 1\,{\rm\AA}^{-1}$, $k_F$ being the Fermi
momentum) but small
compared to the wavelength of light at optical frequencies ($q>>\omega_s/c\sim
0.005\,{\rm\AA}^{-1}$). In a typical
inelastic electron scattering experiment, however, the finite angular
acceptance of the energy-loss spectrometer garantees that the momentum
transfer $q$ be larger than $\omega_s/c$, so that the retarded region of the
surface plasmon dispersion is not observed.\cite{raether2}

\begin{figure}
\includegraphics[width=0.95\linewidth]{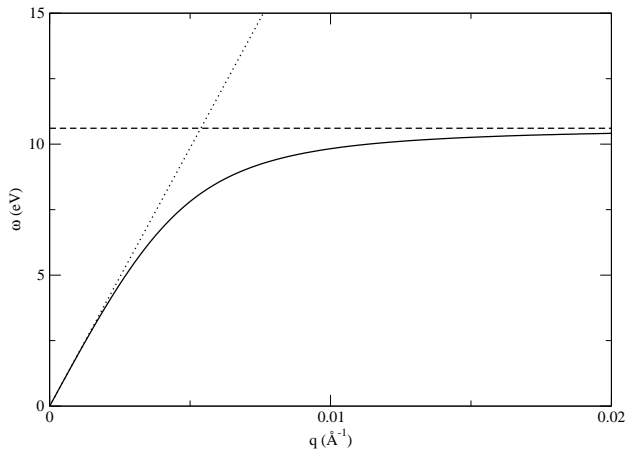}\\
\caption{The solid line represents the dispersion of surface-plasmon polaritons
on a semi-infinite Drude metal with $\omega_p=15\,{\rm eV}$, as obtained from
Eq.~(\ref{retarded4}). In the retarded region ($q<\omega_p/c$), the
surface-plasmon polariton dispersion curve approaches the light line
$\omega=cq$ (dotted line). At short wavelengths ($q>>\omega/c$), the
dispersion curve approaches asymptotically the non-retarded surface-plasmon
frequency $\omega_s=\omega_p/\sqrt 2$ (dashed line).}\label{fig8}
\end{figure}

Considering the full set of Maxwell equations and still assuming that the
wavelength is long enough for a classical description of the metal/vacuum
interface to be justified ($q<<k_F$), one finds that due to retardation the
surface-plasmon condition of Eq.~(\ref{classical}) must be replaced
by\cite{raether2}
\begin{equation}\label{retarded1}
{\epsilon(\omega_s)\over\kappa(\omega_s)}+{1\over\kappa'(\omega_s)}=0,
\end{equation} 
where
\begin{equation}\label{retarded2}
\kappa(\omega)=\sqrt{q^2-\epsilon(\omega){\omega^2\over c^2}}
\end{equation}
and
\begin{equation}\label{retarded3}
\kappa'(\omega)=\sqrt{q^2-{\omega^2\over c^2}}.
\end{equation}
In the case of a semi-infinite Drude metal,
Eqs.~(\ref{retarded1})-(\ref{retarded3}) yield the surface-plasmon dispersion
\begin{equation}\label{retarded4}
\omega^2(q)=\omega_p^2/2+c^2q^2-\sqrt{\omega_p^4/4+c^4q^4},
\end{equation}
or, equivalently,
\begin{equation}
q(\omega)={\omega\over c}\,\sqrt{\omega^2-\omega_p^2\over 2\omega^2-\omega_p^2},
\end{equation}
which we have represented in Fig.~\ref{fig8} by a solid line, together with the
non-retarded
surface-plasmon frequency $\omega_s$ (dashed line) and the light line
$\omega=cq$ (dotted line). In the retarded region, where $q<\omega_s/c$,
surface plasmons couple with the free electromagnetic field, thereby becoming
what is called a surface-plasmon polariton. In the non-retarded limit
($q>>\omega_s/c$), one finds the non-dispersive surface-plasmon frequency
$\omega_s$.

Significant deviations from the classical surface-plasmon dispersion of
Eq.~(\ref{retarded4}) (like the negative dispersion drawn in Fig.~\ref{fig4}),
which are typically observed in electron scattering experiments, are only
present at wave vectors larger than those considered in Fig.~\ref{fig8}. At
${\bf q}$ wave vectors in the range $\omega_s/c<<q<<k_F$, the surface plasmon does
not disperse.

\section{Acoustic surface plasmons}

\begin{figure}
\includegraphics[width=0.75\linewidth]{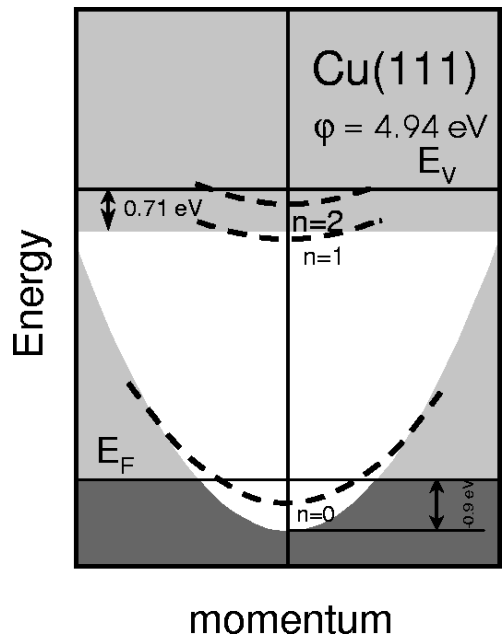}\\
\caption{Schematic representation of the surface band structure on Cu(111) near
the $\bar\Gamma$ point. The shaded region represents the projection of the
bulk bands.}\label{fig10}
\end{figure}

\begin{table}
\caption{Binding energies ($\varepsilon_F^{2D}$) of surface states at the
$\bar\Gamma$ point of Be(0001) and the (111) surfaces of the noble metals Cu,
Ag, and Au. $v_F^{2D}$ and $m^{2D}$ represent the corresponding 2D Fermi velocity
and effective mass, respectively. $v_F^{2D}$ is expressed in units of the Bohr velocity
$v_0=e^2/\hbar$.}
\begin{ruledtabular} \begin{tabular}{lcccccc}
&$\varepsilon_F^{2D}$(eV)&$v_F^{2D}/v_0$&$m^{2D}$\\ \hline
Be(0001)&2.8&0.41&1.18\\
Cu(111)&0.44&0.28&0.42\\
Ag(111)&0.065&0.11&0.44\\
Au(111)&0.48&0.35&0.28\\
\end{tabular}
\end{ruledtabular}\label{table2}
%\vspace{0.2cm}
\end{table}

A variety of metal surfaces, such as Be(0001) and the (111) surfaces of the
noble metals Cu, Ag, and Au, are known to support a partially occupied band of
Shockley surface states with energies near the Fermi level (see
Fig.~\ref{fig10}).\cite{ingles} Since these states are strongly
localized near the surface and disperse with momentum parallel to the surface,
they can be considered to form a quasi 2D surface-state band
with a 2D Fermi energy equal to the surface-state binding energy at the
$\bar\Gamma$ point (see Table~\ref{table2}).

In the absence of the 3D substrate, Shockley surface states would support a 2D
collective oscillation, the energy of this plasmon being given by Eq.~(\ref{2D}) with
$na$
replaced by the 2D density of occupied surface states:
$n^{2D}=\varepsilon_F^{2D}/\pi$. Eq.~(\ref{2D}) shows that at very long
wavelengths plasmons in a 2D
electron gas have low energies; however, they do not affect e-h and phonon
dynamics
near the Fermi level, due to their square-root dependence on the wave vector.
Much more effective than ordinary 2D plasmons in mediating, e.g.,
superconductivity would be the so-called acoustic plasmons with sound-like
long-wavelength dispersion. 

Here we show that in the presence of the 3D substrate the dynamical screening
at the surface provides a mechanism for the existence of a {\it new} acoustic
collective mode, whose energy exhibits a linear dependence on the 2D wave
number.

\subsection{A simple model}

\begin{figure}
\includegraphics[width=0.75\linewidth]{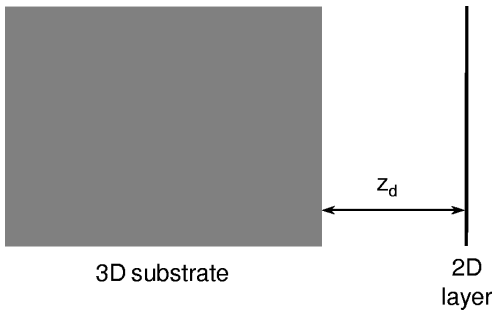}\\
\caption{Surface-state electrons comprise a 2D sheet of interacting free
electrons at $z=z_d$. All other states of the semi-infinite metal comprise a
plane-bounded 3D electron gas at $z\leq 0$. The metal surface is located at
$z=0$.}\label{fig11}
\end{figure}

First of all, we consider a simplified model in which surface-state electrons
comprise a 2D electron gas at $z=z_d$ (see Fig.~\ref{fig11}), while all other
states of the semi-infinite metal comprise a 3D substrate at $z\leq 0$
represented by the Drude dielectric function of Eq.~(\ref{drude}). Within this
model, one finds that both e-h and collective excitations occurring within the 2D
gas can be described with the use
of an effective 2D dielectric function, which in the random-phase
approximation (RPA) takes the form\cite{pitarke1}
\begin{equation}\label{eff}
\epsilon_{eff}^{2D}(q,\omega)=1-\left[{2\pi\over q}+\tilde
W(z_d,z_d;q,\omega)\right]\chi_{2D}^0(q,\omega),
\end{equation}
$\tilde W(z,z';q,\omega)$ being the screened interaction of Eq.~(\ref{w}), and
$\chi_{2D}^0(q,\omega)$ being the noninteracting density-response function of
a 2D electron gas.\cite{stern}

In the absence of the 3D substrate, $\tilde W(z,z';q,\omega)$ is simply zero
and $\epsilon_{eff}^{2D}(q,\omega)$ coincides, therefore, with the RPA dielectric
function
of a 2D electron gas, which in the long-wavelength ($q\to 0$) limit has one
single zero corresponding to collective excitations at $\omega=\omega_{2D}$.

In the presence of a 3D substrate that is {\it spatially separated} from the 2D
sheet
($z_d>0$), the long-wavelength limit of  $\epsilon_{eff}^{2D}(q,\omega)$ has two
zeros. One zero corresponds to a high-frequency oscillation of energy
$\omega^2=\omega_s^2+\omega_{2D}^2$ in which 2D and 3D electrons oscillate in
phase with one another. The other zero corresponds to a low-frequency {\it
acoustic} oscillation in which both 2D and 3D electrons oscillate out of
phase. The energy of this low-frequency mode is found to be of the
form\cite{pitarke1}
\begin{equation}
\omega=\alpha\,v_F^{2D}\,q,
\end{equation}
with $\alpha>1$. For small values of the $z_d$ coordinate ($z_d<<1$), $\alpha\to
1$ and the sound velocity approaches, therefore, the Fermi velocity of the 2D
sheet. For $z_d>>1$, the noninteracting 2D
density-response function takes the Drude form $(1/2\pi)(v_F^{2D}q/\omega)^2$,
and one finds
\begin{equation}
\alpha=\sqrt{2z_d},
\end{equation}
which is the result first obtained by Chaplik in his study of charge-carrier
crystallization in low-density inversion layers.\cite{chaplik}

If the 2D sheet is located inside the 3D substrate ($z\leq 0$), however, the
long-wavelength limit of the effective 2D dielectric function of Eq.~(\ref{eff}) has no
zeros at low energies ($\omega<\omega_s$), due to a complete screening at these
energies of electron-electron interactions within the 2D sheet. This result has
suggested over the years that acoustic plasmons should only exist in the case of {\it
spatially separated}
plasmas, as pointed out by Das Sarma and Madhukar.\cite{sarma}

Nevertheless, Silkin {\it et al.}\cite{silkin} have shown that metal surfaces
where a partially occupied quasi-2D surface-state band {\it coexists} in the
same region of space with the underlying 3D continuum do support a
well-defined acoustic surface plasmon, which could not be explained within the
{\it local} model described above. Furthermore, it has been demonstrated that
it is a combination of the nonlocality of the 3D dynamical screening and the
spill out of the 3D electron density into the vacuum which allows the
formation of 2D electron-density acoustic oscillations at metal surfaces,
since these oscilations would otherwise be completely screened by the
surrounding 3D substrate.\cite{pitarke1}  

\subsection{Full calculation}

In order to achieve a full description of the dynamical response of real metal
surfaces, we first consider a one-dimensional single-particle potential that
describes the main features of the surface band
structure.\cite{chulkov1,chulkov2} We then calculate the eigenfunctions and
eigenvalues of the corresponding hamiltonian, and we evaluate the dynamical
density-response function $\chi^0(z,z';q,\omega)$. Finally, we solve an
integral equation to obtain the RPA interacting density-response function
$\chi(z,z';q,\omega)$. From the knowledge of this function, which describes
bulk and surface states on the same footing, one can obtain within
linear-response theory the electron density induced by an external
perturbation $\phi^{ext}(z;q,\omega)$:
\begin{equation}\label{n}
\delta n(z;q,\omega)=\int dz'\chi(z,z';q,\omega)\phi^{ext}(z';q,\omega),
\end{equation}
and the collective excitations created by an external potential of the form
\begin{equation}\label{ext}
\phi^{ext}(z;q,\omega)=-(2\pi/q){\rm e}^{qz}
\end{equation}
can then be traced to the peaks of the imaginary part of the surface-response
function\cite{liebsch}
\begin{equation}\label{g2}
{\rm Im}\left[g(q,\omega)\right]=\int dz\,{\rm e}^{qz}\,{\rm Im}\left[\delta
n(z;q,\omega)\right].
\end{equation} 

\begin{figure}
\includegraphics[width=0.95\linewidth]{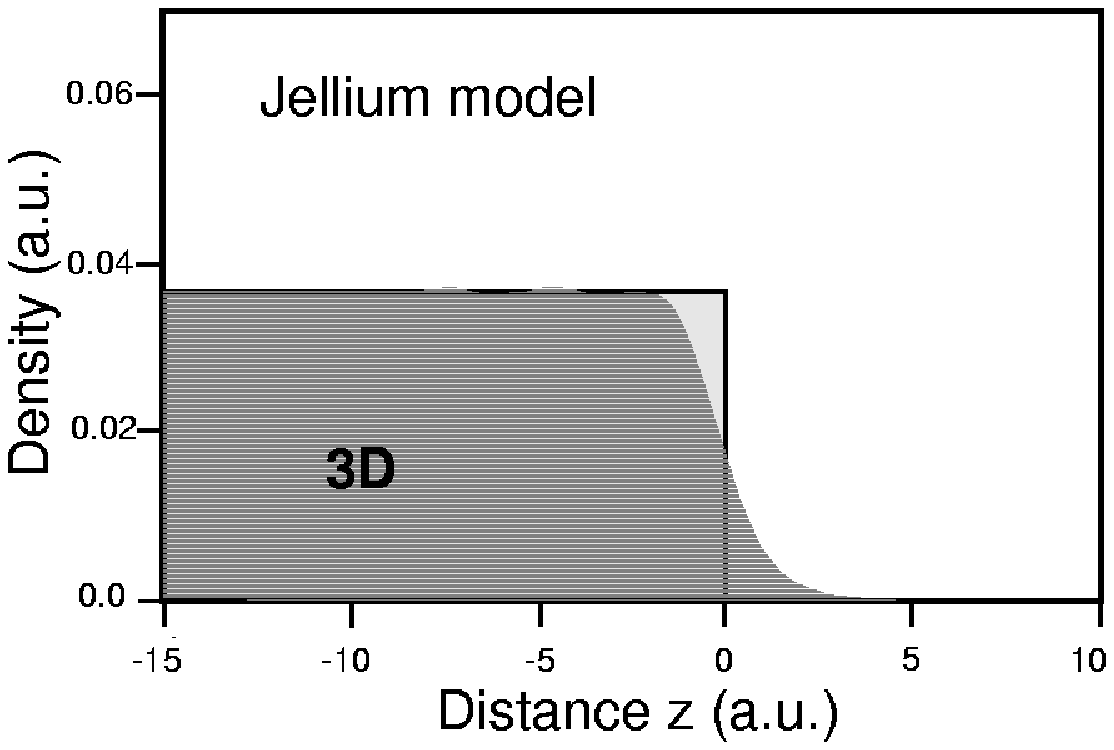}\\
\vspace{1 cm}
\includegraphics[width=0.95\linewidth]{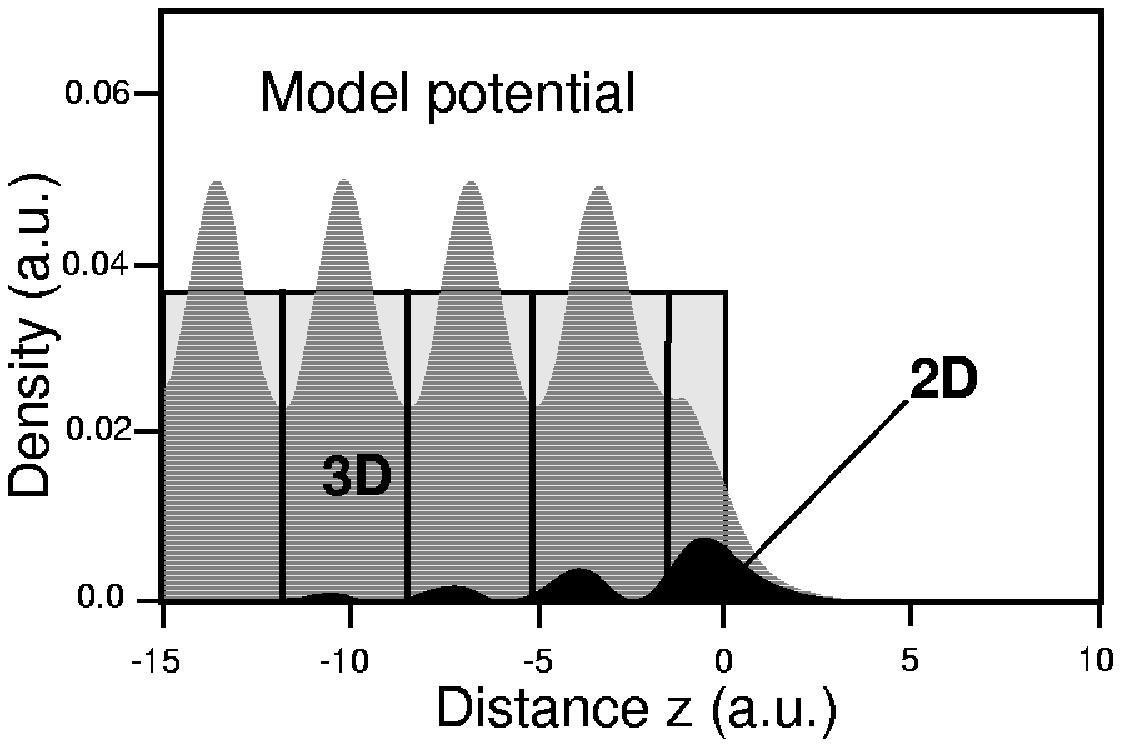}\\
\caption{Unperturbed electron density of the (0001) surface of Be (dark grey
shaded areas), as obtained from a self-consistent jellium density-functional
calculation (top panel) and from the use of a one-dimensional potential that
describes the main features of the surface band structure (bottom panel). The
light grey shaded areas represent the neutralizing uniform positive background.
The vertical lines and black area in the bottom panel, which are absent in the
top panel, represent the atomic positions of the solid and the unperturbed
density of occupied surface states, respectively. The crystal edge ($z=0$) is
chosen to be located half a lattice spacing beyond the last atomic layer, and
$z<0$ corresponds to the interior of the solid.}\label{fig12}
\end{figure}

In the bottom panel of Fig.~\ref{fig12} we show the result that we have obtained
for the unperturbed electron density of the (0001) surface of Be. We see bulk
states, whose total density extends to the interior of the solid at $z<0$, and
surface states which are largely localized near the surface. For comparison, we
have also carried out a self-consistent jellium density-functional calculation of
the electron density (top panel of Fig.~\ref{fig12}), by simply replacing the
ions of the solid by a fixed uniform positive background. Surface states are
absent in this model.

RPA calculations of the imaginary part of the electron density $\delta
n(z;q,\omega)$ induced in a Be(0001) surface by the external potential of
Eq.~(\ref{ext}) were reported in Ref.~\onlinecite{silkin} for $q=0.05\,a_0^{-1}$ ($a_0$ is
the Bohr radius, $a_0=0.529\,{\rm\AA}$) and a wide range of frequencies. It was
demonstrated that this quantity exhibits two
distinct special features, where ${\rm Im}\left[\delta n(q,\omega)\right]$ is
maximum near the surface. The first feature occurs near the surface-plasmon
frequency of valence ($2s^2$) electrons in Be ($\hbar\omega_s=12.8\,{\rm eV}$).
The second feature occurs at $\omega=0.6\,{\rm eV}$, corresponding to a {\it new}
low-energy acoustic collective oscilation, which had been overlooked over the
years.

\begin{figure}
\includegraphics[width=0.9\linewidth]{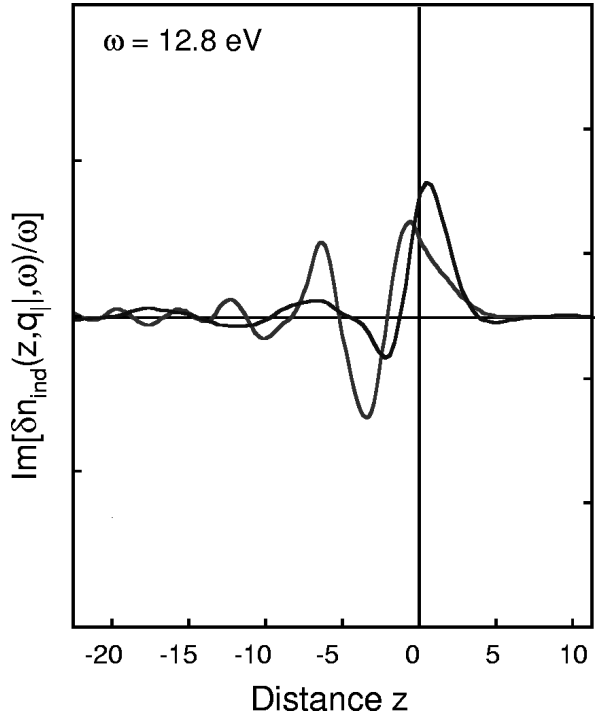}\\
\vspace{1cm}
\includegraphics[width=0.9\linewidth]{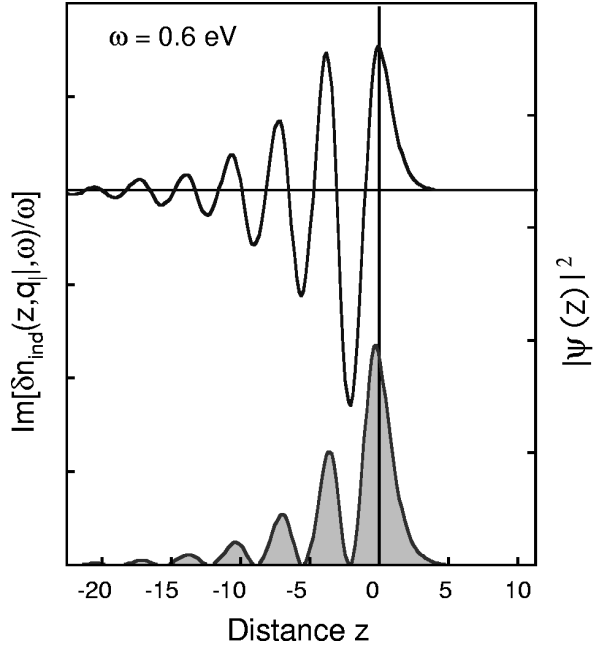}\\
\caption{RPA calculation of the imaginary part of the electron density induced in
the (0001) surface of Be  (black lines), as obtained from Eqs.~(\ref{n}) and (\ref{ext}) for
a wave vector of magnitude $q=0.05\,a_0^{-1}$, as a function of the
$z$ coordinate normal to the surface. The frequency has been chosen to be
$\omega=0.6\,{\rm eV}$ (bottom panel) and $\omega=12.8\,{\rm eV}$ (top panel). The
grey line in the top panel represents the corresponding jellium density-functional
calculation for a semi-infinite free-electron gas. The probability density of the
partially occupied Shockley surface state is represented in the bottom panel by a
shaded area. As in Fig.~\ref{fig12}, the crystal edge ($z=0$) is chosen to be
located half a lattice spacing beyond the last atomic layer, and $z<0$ corresponds
to the interior of the solid.}\label{fig13}
\end{figure}

In Fig.~\ref{fig13} we show by black lines our full RPA calculation of ${\rm
Im}\left[\delta n(z;q,\omega)\right]$ at $q=0.05\,a_0^{-1}$ and the frequencies
$\omega=12.8\,{\rm eV}$ (top panel) and $\omega=0.6\,{\rm eV}$ (bottom panel) at
which surface collective oscillations occur. At the conventional surface-plasmon
frequency $\omega=12.8\,{\rm eV}$ we have also carried out a jellium
density-functional calculation for a semi-infinite free-electron gas, which we
have represented by a grey line in the top panel of Fig.~\ref{fig13}. A comparison
of our band-structure and jellium calculations (black and grey lines) indicates
that the conventional surface plasmon is reasonably well described within a
jellium model of the surface, although Friedel oscillations in the interior of the
solid are considerably more damped in the presence of the actual band structure of
the solid.

At the acoustic surface-plasmon energy, which for $q=0.05\,a_0^{-1}$ is
$\omega=0.6\,{\rm eV}$, a quasi-2D surface-state band in the presence of a 3D
substrate yields the {\it new} feature displayed in the bottom panel of
Fig.~\ref{fig13}. Also shown in this figure (shaded area) is the probability
density of the partially occupied Shockley surface state, clearly indicating that
the low-energy collective excitation at $\omega=0.6\,{\rm eV}$ originates from
this 2D surface-state band. Such a 2D electron gas {\it alone} would only support
a plasmon that for $q=0.05\,a_0^{-1}$ has energy $\omega_{2D}=2.7\,{\rm eV}$, well
above the low-energy excitation that is visible in the bottom panel of
Fig.~\ref{fig13}, and it is only the combination of the strongly localized 2D
surface-state band with 3D bulk states which allows the formation of this new
mode.

\begin{figure}
\includegraphics[width=0.95\linewidth]{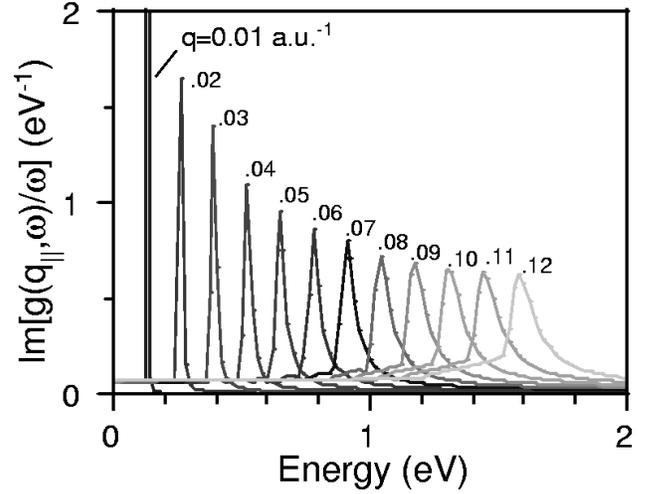}\\
\caption{Energy-loss function ${\rm Im}\left[g(q,\omega)\right]/\omega$ of
Be(0001) versus the energy $\omega$, as obtained from Eq.~(\ref{g2}) for various
values of the wave number $q$, in units of the inverse Bohr radius $a_0^{-1}$.
The peaks are dictated by the corresponding poles of the surface response function
$g(q,\omega)$. In the long-wavelenth ($q\to 0$) limit, $g(q,\omega)$ is simply the
total electron density induced by the potential of Eq.~(\ref{ext}).}\label{fig14}
\end{figure}

Fig.~\ref{fig14} shows the imaginary part of the surface-response function
$g(q,\omega)$ of Be(0001), as obtained from Eq.~(\ref{g2}) for increasing values of
$q$. This figure clearly shows that the excitation spectra is dominated at low
energies by a well-defined {\it acoustic} peak with {\it linear} dispersion, the
sound velocity being at long wavelengths very close to the 2D Fermi velocity
$v_F^{2D}$ (see Table~\ref{table2}).

\begin{figure}
\includegraphics[width=0.75\linewidth]{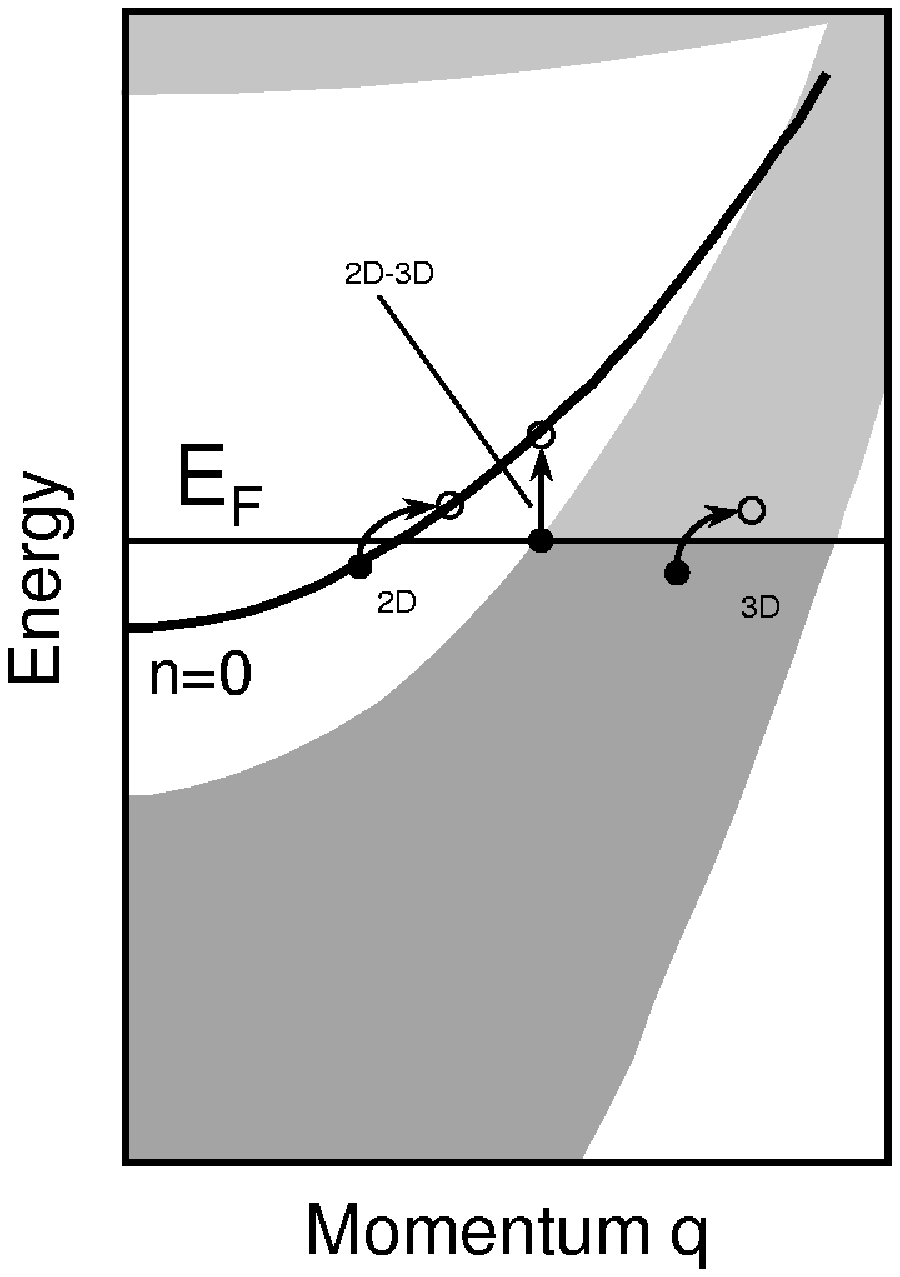}\\
\vspace{1cm}
\includegraphics[width=0.95\linewidth]{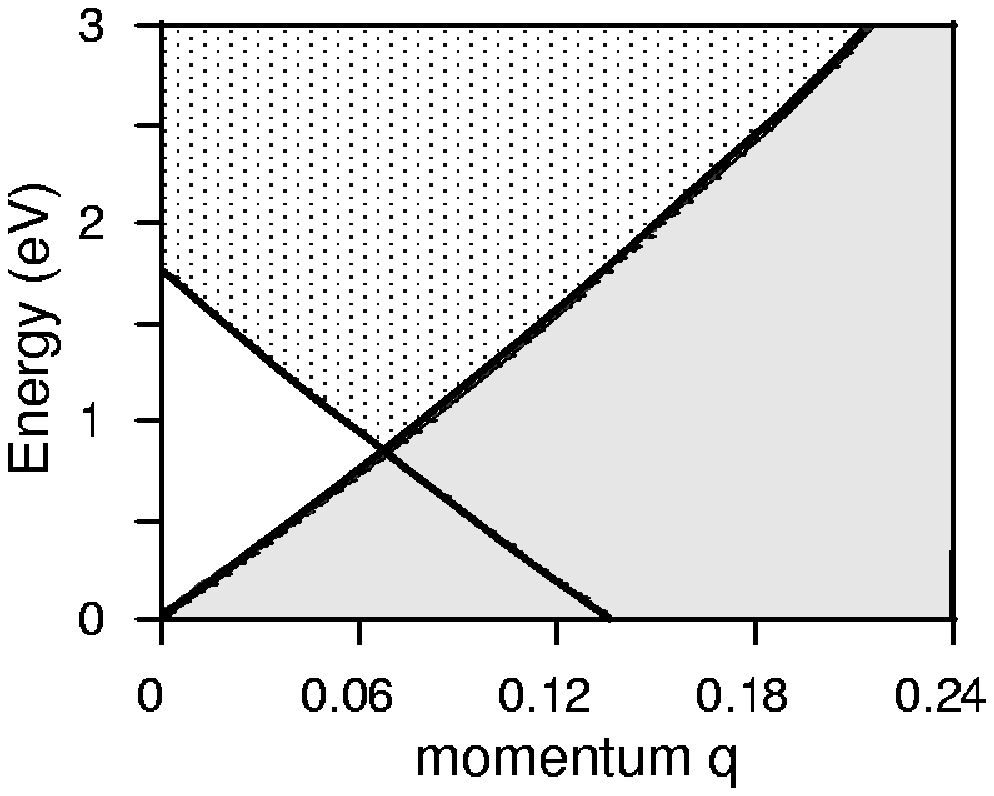}\\
\caption{Top panel: Schematic representation of the projection of the bulk bands
onto a solid surface that supports a partially occupied Shockley surface-state
band. The energy of occupied and unoccupied states is displayed as a function of
the momentum parallel to the surface. The solid line represents a Shockley surface
state. Dark and light shaded areas represent occupied and unoccupied bulk states,
respectively. The white area represents the band gap. Electron-hole pair
excitations are represented by arrows, depending on whether they correspond to
transitions within the surface-state band (2D), transitions within the bulk bands
(3D), or transitions from bulk to surface states (2D-3D). Bottom panel: Dispersion
of the acoustic surface collective excitation of Be(0001), as derived from the
peaks that are visible in Fig.~\ref{fig14} (thick solid line). This curve stays
just over the upper edge $\omega_u^{2D}=v_F^{2D}q+q^2/2$ of the 2D e-h pair
continuum (shaded area). Momentum and energy conservation prevent 2D e-h pairs
from being produced for energies above $\omega_u^{2D}$. The area below the thin
solid line represents the region of momentum space where transitions from 3D to 2D
states cannot occur. Well-defined acoustic plasmons are expected to occur at wave
vectors with magnitude smaller than $q\sim 0.06\,a_0^{-1}$.}\label{fig15}
\end{figure}

In the bottom panel of Fig.~\ref{fig15} we show the energy of the acosutic surface
plasmon in Be(0001) versus the wave number $q$ (thick solid line), as derived from
the maxima of our calculated surface-loss function ${\rm
Im}\left[g(q,\omega)\right]$ of Fig.~\ref{fig14}. The plasmon energy of electrons
in an isolated 2D electron gas would exhibit at long wavelengths a square-root
dispersion with the wave number $q$ [see Eq.~(\ref{2D})]. However, the combination
of a 2D surface-state band with the underlying 3D system yields a {\it new}
distinct mode whose energy lies just above the upper edge
$\omega_u^{2D}=v_F^{2D}q+q^2/2$ of the continuum of 2D e-h pair excitations
(shaded area), where momentum and energy conservation allows e-h pairs to be
created within the 2D electron gas.

For a well-defined acoustic surface plasmon to occur, it must exist for wave
vectors ${\bf q}$ and energies $\omega$ where decay cannot occur by exciting e-h
pairs in the medium. Electron-hole pairs can be excited either within the 2D
surface-state band, or within the 3D continuum of bulk states, or by promoting an
electron from an occupied bulk state to an unoccupied 2D surface state. These
three mechanisms for the production of e-h pairs at low energies are illustrated
in Fig.~\ref{fig15}. (i) Since the energy of acoustic surface plasmons always lies
just above the upper edge $\omega_u^{2D}$ of the 2D e-h pair continuum (shaded
area in the bottom panel of Fig.~\ref{fig15}), they cannot possibly decay by
creating e-h pairs within the 2D band. (ii) The 3D Fermi velocity $v_F^{3D}$ is typically
larger than the Fermi velocity $v_F^{2D}$ of the 2D surface-state band. This means
that acoustic surface plasmons can decay by exciting e-h pairs within the 3D
continuum of bulk states. However, at the low energies involved the probability for this
process to occur is small.\cite{pitarke1} (iii) An inspection of the upper panel of
Fig.~\ref{fig15} shows that, due to the presence of the band gap, for optical ($q=0$)
transitions to occur from an occupied 3D bulk state to an unoccupied 2D surface state
a minimum energy is required, which decreases as the momentum transfer $q$
increases.

The region of momentum space where transitions from 3D to 2D states cannot occur
corresponds to the area below the thin solid line in the bottom panel of
Fig.~\ref{fig15}. This figure shows that at long wavelengths with $q<0.06\,a_0^{-1}$
acoustic surface plasmons can decay by neither exciting 2D e-h pairs nor exciting
3D-2D e-h pairs, which results in a very well-defined collective excitation (see
Fig.~\ref{fig14}). At shorter wavelengths ($q>0.06\,a_0^{-1}$), the promotion of electrons
from occupied 3D bulk states to unoccupied 2D surface states becomes
possible and the corresponding plasmon peak broadenes considerably.

\subsection{Excitation of acoustic surface plasmons}

We close this paper by discussing whether acoustic surface plasmons can be
observed. As in the case of conventional surface plasmons, acoustic surface
plasmons should be expected to be excited by either electrons or light. Here we
focus on a possible mechanism that would lead to the excitation of acoustic
surface plasmons by light in, e.g., vicinal surfaces with high indices.

\begin{figure}
\includegraphics[width=0.75\linewidth]{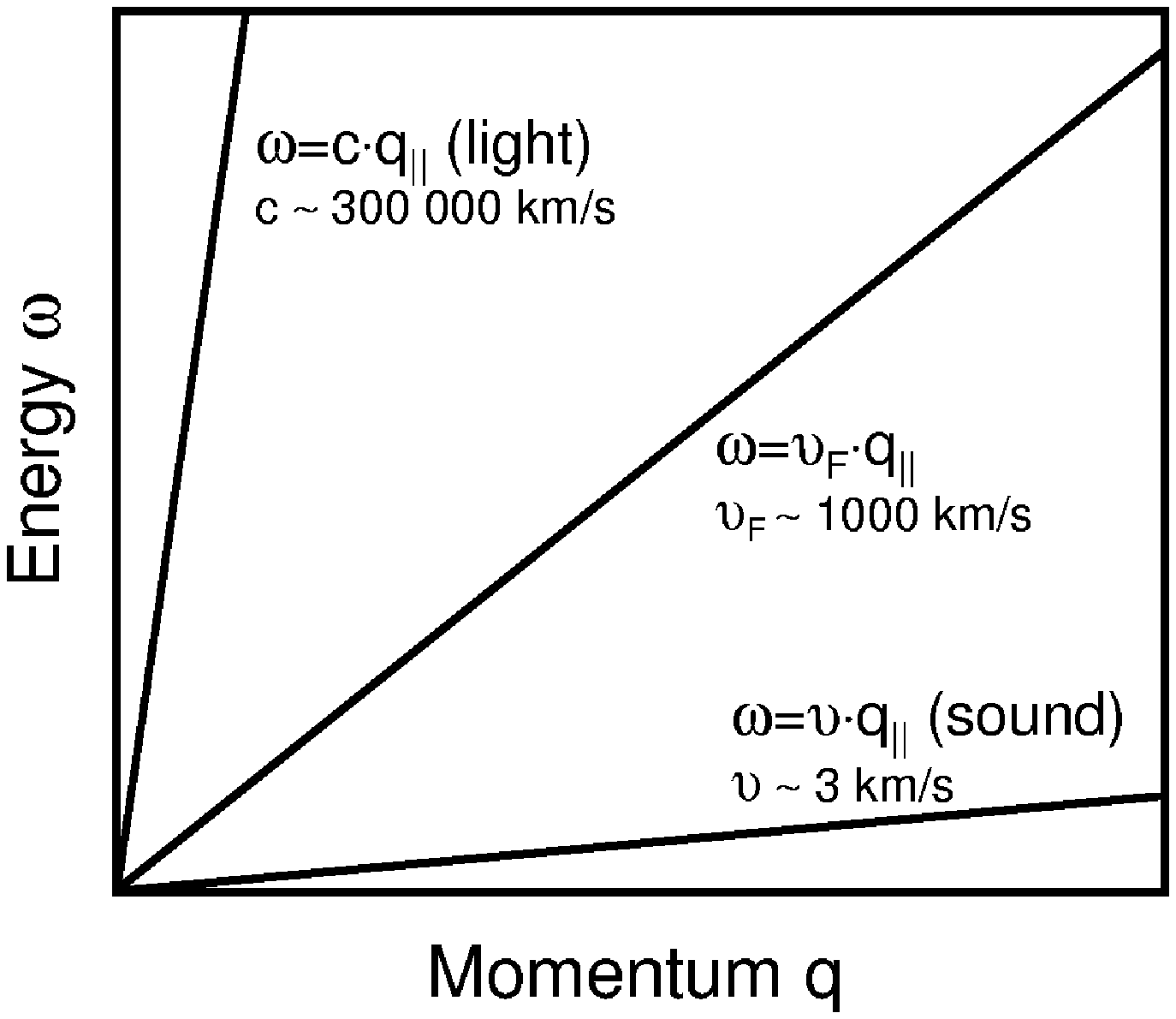}\\
\vspace{1 cm}
\includegraphics[width=0.75\linewidth]{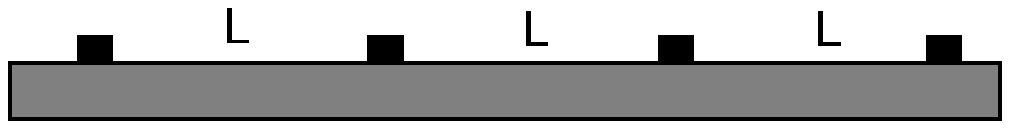}\\
\caption{Top panel: Schematic (out of scale) representation of typical energy
dispersions of acoustic surface plasmons, acoustic phonons, and free-space
electromagnetic radiation. In the actual scale, the dispersion lines of free light
and acoustic phonons should be very close to the vertical and horizontal axes,
respectively. Bottom panel: Periodic grating of constant $L$. The grating periodic
structure can provide the impigning free electromagnetic radiation with additional
momentum $2\pi/L$.}\label{fig16}
\end{figure}

The top panel of Fig.~\ref{fig16} exhibits the energy dispersion of acoustic
surface plasmons at low wave vectors. Also shown in this figure are the
light line $\omega=cq$ and, for comparison, a typical energy dispersion of
acoustic phonons. The sound velocity of acoustic surface plasmons, which is close
to the Fermi velocity of the 2D surface-state band, is typically a few orders of
magnitude larger than the sound velocity of acoustic phonons in metals but still
about two orders of magnitude smaller than the velocity of light.

The acoustic surface plasmon dispersion curve is well below the dispersion curve
of free-space electromagnetic radiation. Hence, there is, in principle, no way
that incident light can provide an ideal surface with the correct amount of
momentum and energy for the excitation of an acoustic surface plasmon. As in the
case of conventional surface plasmons, however, a periodic corrugation or grating
in the metal surface should be able to provide the missing momentum. 

If light hits a surface with a periodic corrugation, the grating (see bottom panel
of Fig.~\ref{fig16}) can provide the impigning free electromagnetic waves with
additional momentum arising from the grating periodic structure. If free
electromagnetic radiation hits the grating at an angle $\theta$, its wave vector
along the grating surface has magnitude
\begin{equation}\label{grat}
q={\omega\over c}\,\sin\theta\pm{2\pi\over L}\,n,
\end{equation}
where $L$ represents the grating constant, and $n=1,2,\dots$. Hence, the linear
(nearly vertical) dispersion relation of free light changes into a set of parallel straight
lines, which can match the acoustic-plasmon dispersion relation.

For a well-defined acoustic surface plasmon in Be(0001) to be observed, the
wave number $q$ needs to be smaller than $q\sim 0.06\,a_0^{-1}$.
For $q=0.05\,a_0^{-1}$, Eq.~(\ref{grat}) with $n=1$ yields a grating constant
$L=66\,{\rm\AA}$. Acoustic surface plasmons of energy $\omega\sim 0.6\,{\rm eV}$
could be excited in this way. Although a grating period of a few nanometers
sounds unrealistic with present technology, the possible control of vicinal
surfaces with high indices could provide appropriate grating periods in the
near future. Alternatively, acoustic surface plasmons could be observed
with the use of high-resolution electron energy-loss spectroscopy (EELS) under
grazing incidence. 
 
\acknowledgments

Partial support by the University of the Basque Country, the Basque
Unibertsitate eta Ikerketa Saila, the Spanish Ministerio de Educaci\'on y
Cultura, and the Max Planck Research Award Funds is gratefully acknowledged.


\begin{thebibliography}{100}
\bibitem{optical} The valence electron density $n$ is usually characterized by 
the density parameter $r_s=(3/4\pi n)^3/a_0$ ($a_0$ is
the Bohr radius, $a_0=0.529\,{\rm\AA}$). In metals ($2<r_s<6$), plasmons have
energies in the range $5{\rm eV}<\omega_p<20{\rm eV}$ and frequencies that 
lie, therefore, in the optical regime.
\bibitem{long} Throughout this paper, long wavelengths are associated with low wave vectors.
\bibitem{note1} The sound velocity of this acoustic mode is, however, close to
the Fermi velocity of the 2D surface-state band, which is typically a few
orders of magnitude larger than the sound velocity of acoustic phonons in
metals but still about two orders of magnitude smaller than the velocity of
light.
\bibitem{notenew} This approximation is valid for heavy charged particles,
e.g., ions, and also for swift electrons moving with nonrelativistic velocities
that are large compared to the velocity of target electrons. In the case of
electrons, $Z_1=-1$.
\end{thebibliography}
\end{document}